\documentclass[aps,pre,twocolumn,superscriptaddress,showpacs,showkeys ]{revtex4-1}
\usepackage{color}
\usepackage{graphicx}
\usepackage{hyperref}
\usepackage{amsmath,amssymb}
\usepackage{epstopdf}
\usepackage{subcaption}
\usepackage{dcolumn}   
\usepackage{bm}        
\usepackage{verbatim}   

\begin{document}

\title{Edwards-Wilkinson Depinning Transition in Random Coulomb Potential Background}

\author{N. Valizadeh}
\affiliation{Physics Department, K. N. Toosi University of Technology, Tehran, Iran}

\author{M. Samadpour}
\affiliation{Physics Department, K. N. Toosi University of Technology, Tehran, Iran}

\author{H. Hamzehpour}
\affiliation{Physics Department, K. N. Toosi University of Technology, Tehran, Iran}

\author{M. N. Najafi}
\affiliation{Department of Physics, University of Mohaghegh Ardabili, P.O. Box 179, Ardabil, Iran}
\email{morteza.nattagh@gmail.com}

\begin{abstract}
The quenched Edwards-Wilkinson (QEW) growth of $1+1$ interface is considered in the background of the correlated random noise. We use random Coulomb potential as the background long-range correlated noise. A depinning transition is observed in a critical driving force $F_c\approx 0.37$ in the vicinity of which the final velocity of the interface varies linearly with time. Our data collapse analysis for the velocity shows a crossover time $t^*$ at which the velocity is size independent. Based on a two-variable scaling analysis, we extract the exponents, which are different from all universality classes we are aware of. Especially noting that the dynamic and roughness exponents are $z_w=1.55\pm 0.05$, and $\alpha_w=1.05\pm 0.05$ at the criticality, we conclude that the system is different from both EW and KPZ universality classes. Our analysis shows therefore that making the noise long-range-correlated, drives the system out of EW universality class. The simulations on the tilted lattice shows that the non-linearity term ($\lambda$ term in the KPZ equations) goes to zero in the thermodynamic limit.
\end{abstract}

\pacs{05., 05.20.-y, 05.10.Ln, 05.45.Df}
\keywords{invasion percolation, fluid dynamics, critical exponents}

\maketitle

\section{Introduction}
The growth of rough interfaces in random media has absorbed much attentions due to its vast application in the fluid dynamics in porous media~\cite{pismen2006patterns,collet2014instabilities,cross1993pattern}, fluid-fluid displacement~\cite{rubio1989self,horvath1991anomalous}, fire front motion~\cite{zhang1992modeling} and motion of flux lines in superconductors~\cite{blatter1994vortices}. Magnetic domains~\cite{allenspach1990magnetic} and cell membranes~\cite{moglia2016pinning} are other examples of the growth of rough interfaces in random media. An important large class of interface dynamics are the ones which are pushed, and at the same time get pinned at random obstacles already present in the host media. The stochasitisity of the dynamics of these interfaces is due to the latter, which originates from the stochasticity of the obstacles, i.e. their size, permeability, etc. Generally the properties of the interfaces depend on two ingredients~\cite{bru2004pinning}: the dynamical laws governing the interfaces and the pattern of the quenched disorder present in the host media. The effect of the governing laws has been vastly studied in the literature~\cite{barabasi1995fractal}.  The depinning transition phenomenon is one of the important observations in this field, which is defined as a point that the system changes behavior from being pinned to obstacles (which is identified by a zero interface velocity at long times), to a moving phase (non-zero interface velocity at long times). This second order transition, for which there are many experimental and theoretical evidences, is due to the non-linearity of these systems, and the competition between the driving force and the resistance force due to the obstacles~\cite{amaral1995scaling}. In the theoretical side, there are two general approaches for modeling these systems: continuous and discrete models which are classified based on the critical exponents in the vicinity of the transition. quenched Edwards-Wilkinson (QEW) and Kardar-Parizi-Zhang (KPZ) classes are the most famous continuous models, and the directed percolation depinning (DPD) class is one of the most important discrete models. For a good classification of the models see~\cite{barabasi1995fractal,amaral1995scaling}. Many aspects and applications of these models were studied in the literature  ~\cite{barabasi1995fractal,bru2004pinning}, ~\cite{bonachela2011universality,azimzade2019effect}. Despite of this huge literature, a little work has been done on the effect of the type of the quenched noise (mainly correlated vs uncorrelated) in the host media.
 In most applications, the authors concentrated on the quenched disorders which is realized by the uncorrelated noises in a given range. Actually the nature may behave in a more complicated way, i.e. the disorder can be correlated in various ways which should be realized by the models which contain the key parameters as the original system. Here we focus on the opposite limit of uncorrelated noises, which is two-dimensional long-range correlated disordered systems. We use the Random Coulomb Potential (RCP) to model the quenched noise, which is a well-known model in the statistical mechanics. Many properties of this model are known, for example it is quite similar to the Edwards-Wilkinson in the stationary phase ~\cite{cheraghalizadeh2018gaussian2,cheraghalizadeh2018gaussian}, which itself is $c=1$ conformal field theory~\cite{francesco2012conformal} in two dimensions,  and Coulomb gas~\cite{cardy2005sle}. In the global point of view, the interfaces of this model is described by Schramm-Lowener evolution with the diffusivity parameter $\kappa=4$~\cite{cardy2005sle}. The host media which is governed by RCP is long-range, since the correlation of fields is logarithmic (equivalent to the zero roughness exponent in EW model) and therefore the corresponding noise is quite long-range~\cite{stanley1993long,knackstedt2000invasion}. Previously some dynamical models combined with random coulomb potential model, like the percolation~\cite{cheraghalizadeh2018gaussian} and the Ising~\cite{cheraghalizadeh2018gaussian2}.\\
In this paper we consider the depinning transition governed by the QEW model on top of a media for which the correlation of the noise is controlled by RCP. We find that the properties of the depinning transition changes significantly by introducing the RCP correlations. Especially we see a new type two-parameter scaling relations around the critical driving force $F=F_c$.    \\

The paper is organized as follows: in the next section we explore the main properties of the depinning transitions. The general set up of the RCP noise in pour paper is presented in SEC.~\ref{RoughSurfaces}. Section~\ref{results} is devoted to the presentation of our results. We close the paper by a conclusion.

\section{General Properties of Driven Interfaces}
We study the driven interface dynamics in two-dimensional random media as its background, represented by stochastic obstacles, whose positions and strengths are correlated. The problem is $2+1$ dimensional, where two spatial components are $(x,y)$, and $1$ stands for the time $t$. We especially concentrate on the motion pattern of one-dimensional interface $y=h(x,t)$, the border between ``dry'' and ``wet'' phases, in the presence of ``quenched noise'' $\eta(x,y)$ distributed over the space, and also driving force (here represented by $F$). For small enough driving forces the interface is shown to be in the ``pinned'' phase in which the interface stops, more precisely $\bar{h}(t)\equiv \sum_{x=1}^L h(x,t)$ (where $L$ is lattice size) vanishes at large enough time due to the obstacles, meaning that the disordered resistance force is statistically more effective than the driving force. While for, i.e. large enough $F$s the interfaces are in the ``moving'' phase. There is a critical value of $F$, represented by $F_c$ where the interface undergoes a transition from the pinned phase to the moving phase, named as the depinning transition. $\bar{v}_{\infty}\equiv \lim_{t\rightarrow \infty}\frac{d}{dt}\bar{h}$ plays the role of the order parameter of this transition, i.e. $\bar{v}_{\infty}=0$ for the pinned phase, and $\bar{v}_{\infty}>0$ for the moving phase. When $\eta(x,h(x,t))$ is an uncorrelated quenched random noise, and the governing equation is the Edwards-Wilkinson (EW), not surprisingly the underlying interface at $F_c$ becomes a self-similar (more precisely self-affine) extended object with critical properties similar to one-dimensional EW universality class. In this case, many statistical observables display scaling behaviors. The example is the interface width, which  has become a standard tool in the study of growing surfaces for various theoretical and experimental models of growing interface. It characterizes the roughness of the interface, being defined by the fluctuations of the height field
It is defined by
\begin{equation}
	w^2=\left\langle \overline{\left(h(\textbf{r})-\bar{h} \right)^2} \right\rangle,
\end{equation}
where the over line represents the spatial average $\overline{O}\equiv \frac{1}{L}\sum_i O(x=i)$, the $\left\langle \cdots \right\rangle $ is the ensemble average. For the EW model it is well-known that there is a cross over time scale $t_X$. For $t\leq t_X$ the width increases as a power of time $w(L,t)\sim t^{\beta_{w}}$, where the exponent $\beta_{w}$ is call the growth exponent, and characterizes the time-dependent dynamics of the roughening process. The power-law increase of width does not continue indefinitely, but is followed by a saturation regime (for $t\geq t_X$) during which the width reaches a saturation value, $w_{\text{sat}}$. Indeed $w_{\text{sat}}$ itself scales with $L^{\alpha_{w}}$, the exponent $\alpha_{w}$ being called the roughness exponent, is a second critical exponent that characterizes the model. The time scale $t_X$ depends on $L$ in a power-law fashion $\sim L^{z_{w}}$, where the dynamic exponent $z_w$ is equal to $\frac{\beta_w}{\alpha_w}$. These relations are summarized in a famous scaling relation for the interface width, $w$, as well as some other statistical quantities~\cite{barabasi1995fractal,kondev2000nonlinear,nezhadhaghighi2011contour,najafi2017scale}
\begin{equation}
	w(L,t)=L^{\alpha_w}F_w\left(\frac{t}{L^{z_w}}\right) =t^{\frac{\alpha_w}{z_w}}G_w\left(\frac{t}{L^{z_w}}\right)
	\label{Eq:roughness-FSS}
\end{equation}
from which the exponents can be extracted. The functions $F_w(x)$ and $G_w(x)=x^{-\frac{\alpha_w}{z_w}}F(x)$ (showing that $z_W=\frac{\alpha_w}{\beta_w}$) are some universal functions with the asymptotic behavior $F_w(x)=\left\lbrace\begin{matrix}
	x^{\beta_w} & x\ll 1\\
	const. & x\gg 1
\end{matrix} \right. $. \\
For $1+1$ EW the exponents are $\alpha_w=0.5$, $\beta_w=0.25$, and $z_w=2$, whereas for QEW we have $\alpha_w=0.92(4)$, $\beta_w=0.85(3)$, and $z_w=1.08(1)$.\\ 

For the QEW class at $F=F_c$ the average velocity $\left\langle v(t)\right\rangle $ decreases with time in a power-law fashion~\cite{moglia2016pinning}
\begin{equation}
\left\langle v(t)\right\rangle\sim t^{-q}
\label{Eq:dynamicalPowerLow}
\end{equation}
whereas $\left\langle v(t)\right\rangle\sim e^{-t/\xi_F}$ for $F<F_c$, where $\xi_F\sim |F-F_c|^{-\nu}$ and $\nu$ is called correlation length exponent. In the vicinity of $F_c$, we have also
\begin{equation}
\bar{v}_{\infty}\sim f^{\theta}, \ \ \ f\equiv\frac{F-F_c}{F_c}
\label{Eq:v-theta}
\end{equation}
where $\theta>0$ is the velocity exponent. In fact close to this transition, some parts of the interface is growing, and some parts are \textit{pinned}, forming pinning paths, and the growth occurs by propagation of these growing regions. Taking into account that the characteristic time required for this propagation is $t_X$ (since it is the time required for correlations to propagate across the system), and the typical advance for each movement (from one blocking path to the other) is $w_{\text{sat}}$, one obtains~\cite{amaral1995scaling}
\begin{equation}
\bar{v}_{\infty}\propto \frac{w_{\text{sat}}}{t_X}\propto \zeta^{\alpha}/\zeta^{z}\propto f^{\nu(z-\alpha)}
\end{equation} 
giving us the hyperscaling relation $\theta=\nu(z-\alpha)$.\\
There are many experiments to realize the driven interfaces, like fluid-fluid displacement~\cite{stokes1988interface,rubio1989self,horvath1991dynamic}, imbibition of cofee in paper towels~\cite{buldyrev1992anomalous,amaral1994new,amaral1995avalanches}, which reported scattered exponents for the critical driven interfaces~\cite{amaral1995scaling}. Also many effects in this field has been studied, like
$1/f$ noise in driven interfaces~\cite{krug19911}, anomalous noise in driven interfaces~\cite{horvath1991anomalous}.
Recently it was conjectured by Grassberger that critically pinned interfaces in 2-dimensional isotropic random media with short range correlations are always in the universality class of ordinary percolation~\cite{grassberger2018universality}.
There is a clever method to distinguish the universality classes of the driven interfaces, especially the KPZ universality class. If we start from a tilted initial configuration, i.e. $h_0=mx$, then the final velocity behaves like
\begin{equation}
\bar{v}_{\infty}(m)=\bar{v}_{\infty}(m=0)+\lambda m^2
\label{Eq:tilted}
\end{equation}
where $\bar{v}_{\infty}(m=0)=\bar{v}_{\infty}\propto f^{\theta}$ scales with $f$ as before, and $\lambda\sim f^{\phi}$, where $\phi<0$ for KPZ universality class, and $\phi>0$ for EW universality class. We see that
\begin{equation}
	\bar{v}_{\infty}(m)\propto f^{\theta}+a f^{\phi} m^2
	\label{Eq:tilted2}
\end{equation}
where $a$ is an unimportant constant. Given that $\lambda$ is proportional to the non-linearity coefficient in the KPZ model, one can see that the former is equivalent to large non-linearity term in the transition point governed by the KPZ class.

\section{two-dimensional random coulomb potential noise; \\ Our Model}\label{RoughSurfaces}
\label{sec:random coulomb potential}

In this section we consider a 2D host system with correlated scale-invariant disorder. Many random systems are described in terms of/mapped to the Random Coulomb Potential (RCP), ranging from the free Bosonic system, to Edwards-Wilkinson (EW) model of surface growth process~\cite{hosseinabadi2013universality}, inverse turbulence cascades~\cite{bernard2007inverse}, electric field of random charged noise~\cite{cheraghalizadeh2018gaussian,cheraghalizadeh2018gaussian2}. Here we consider a different realization of RCP is the Poisson equation in the background of white-noise charge disorders, which itself is mapped to EW model in the stationary state. We construct a correlated random (quenched) noise system, through which the driven interface move. The governing equations of the interface dynamics is considered to be the one for QEW equations. We can imagine of this problem as \textit{the coupling of the driven interface problem with the random coulomb potential model}, or the critical phenomena on the fractal systems~\cite{gefen1980critical}. This concept can be extended to dilute systems that are fractal in some limits~\cite{cheraghalizadeh2017mapping,najafi2016bak,najafi2016monte,najafi2016water,najafi2018coupling}. \\
Before describing the problem in this type of media, let us first briefly introduce our method of generating RCPs. The EW model in the stationary state becomes RCP which is generated by the following equation for the height field $V(\textbf{r})$:
\begin{equation}
\partial_t V(\vec{r},t)=\nabla^2V(\vec{r},t)+\eta(\vec{r},t),
\label{Eq:random coulomb potential}
\end{equation}
in which $\eta(\vec{r},t)$ is a space-time white noise with the properties $\left\langle \eta(\vec{r},t)\right\rangle = 0 $ and $\left\langle \eta(\vec{r},t)\eta(\vec{r}',t')\right\rangle = \zeta \delta^3(\vec{r}-\vec{r}')\delta(t-t') $ and $\zeta$ is the strength of the noise. $V(\vec{r},t)$ can be served as the electrostatic potential in our paper, once it becomes $t$-independent (the stationary state of EW, in which on average $\partial_t \left\langle V\right\rangle =0$). In this situation, one may replace the equation with a time-independent equation:
\begin{equation}
\nabla^2V(\vec{r})=-\rho(\vec{r}),
\label{Eq:Poisson}
\end{equation}
which is the Poisson equation with the dielectric constant $\epsilon= 1$. In this equation $\rho(\vec{r})$ is the spatial white noise with the normal distribution and the properties $\left\langle \rho(\vec{r})\right\rangle = 0 $ and $\left\langle \rho(\vec{r})\rho(\vec{r}')\right\rangle = (n_ia)^2 \delta^3(\vec{r}-\vec{r}')$, $n_i$ is the total density of Coulomb disorder and $a$ is the lattice constant. This connection can be confirmed from both numerical and analytical levels. In the theoretical level, if one takes a look at the probability measure of the Eq.~\ref{Eq:Poisson}, finds that it is exactly the same as the probability measure of the EW in the stationary state (see Ref.~\cite{barabasi1995fractal}). On the numerical level also, we have seen that all of the statistical observables are the same. For example, the two-point correlations are logarithmic, the fractal dimension of iso-height contours are $1.5$, the critical exponent of the gyration radius is $3.0$, and the critical exponent of the loop lengths is $7/3$ in accordance with RCP~\cite{kondev1995geometrical}. It is well-known that this model in the scaling limit belongs is described by Gaussian distribution function (RCP) which is $c=1$ conformal filed theory~\cite{francesco1996conformal}. It is also known that the contour lines of this model are described by the Schramm-Loewner evolution (SLE) theory with the diffusivity parameter $\kappa=4$~\cite{cardy2005sle}, which is understood in terms of the general CFT/SLE correspondence with the relation $c=(6-\kappa)(3\kappa-8)/(2\kappa)$. The fractal dimension of the contour loops $D_f^{\text{RCP}}=\frac{3}{2}$ which is also compatible with the relation $D_f=1+\frac{\kappa}{8}$. \\

The probability distribution function of these fields transform under $\textbf{r}\rightarrow b\textbf{r}$ as follows:
scaling law
\begin{eqnarray}\label{scaleinvariance}
V(b \mathbf{r}) \stackrel{d}{=} b ^{\alpha_{\text{RCP}}} V(\mathbf{r}),
\end{eqnarray}
where the parameter $\alpha_{\text{GFF}}$ is \textit{roughness} exponent or the \textit{Hurst} exponent of the RCP and $b$ is a scaling factor and the symbol $\stackrel{d}{=}$ means the equality of the distributions. Let us denote the Fourier transform of $V(\textbf{r})$ by $V(\textbf{q})$. The distribution of this system, like a wide variety of random fields, is Gaussian~\cite{kondev1995geometrical}. Various correlation functions (e.g. $C(r) \equiv \langle \left[ V(\mathbf{r}+\mathbf{r_0})-V(\mathbf{r_0}) \right]^2 \rangle$), and the height total variance show power-law behavior~\cite{barabasi1995fractal}, defining the roughness exponent of which are $\alpha=0$. It is shown that~\cite{kondev2000nonlinear,adler1981geometry} the probability distribution functions of this RCP noise is Gaussian
\begin{eqnarray}
P( V ) \equiv \frac{1}{\sigma\sqrt{2\pi}}e^{-\frac{V^2}{2\sigma^2}},
\end{eqnarray}
where $\sigma$ is the standard deviation. In addition the contour loop ensemble can be characterized through the loop correlation function $G(\mathbf{r})=G(r)$ ($r\equiv |\textbf{r}|$) which is the probability measure of how likely the two points separated by the distance $r$ lie on the same contour. For large distances this function scales with $r$ as~\cite{kondev1995geometrical}
\begin{eqnarray}\label{loop correlation function}
G(r) \sim \frac{1}{r^{2x_l}},
\end{eqnarray}
where $x_l$ is the loop correlation exponent. It is believed that the exponent $x_l$ is super-universal, i.e. for all the known mono-fractal Gaussian random fields in two dimensions this exponent is equal to $\frac{1}{2}$~\cite{kondev1995geometrical,kondev2000nonlinear}. This model belongs to $c=1$ CFT and also SLE$_{4}$.\\

We study the flow of a fluid in a two-dimensional random media as its background, represented by stochastic obstacles, whose positions and strengths are correlated as explained above ($V(\textbf{r})$ in Eq.~\ref{Eq:Poisson}). The dynamics is governed by the QEW continuous growth equation. The system is $2+1$ dimensional, where two spatial components are $(x,y)$, and $1$ stands for the time $t$. We especially concentrate on the motion pattern of one-dimensional interface $y=h(x,t)$, the border between ``dry'' and ``wet'' phases. The motion of $h(x,t)$ is governed by the QEW equation
\begin{equation}
\frac{\partial h(x,t)}{\partial t}=\upsilon\nabla^{2}h+F+V(x,h(x,t)).
\label{Eq:EW}
\end{equation}
where $\upsilon$ is surface tension, $F$ is the driving force as defined above and $\eta(x,y)$ is the ``quenched noise'' distributed over the space. Here we model the noise by random coulomb potential $\eta(x,y)\equiv V_{\text{normalized}}(x,y)$, where $V_{\text{normalized}}(x,y)$ is the normalized $V(x,y)$, so that it takes its values from $[-1,1]$. 

\section{Results}\label{results}
We consider the time evolution of a rough ($1+1$)-dimensional interface described by the vertical coordinate $y=h(x,t)$ (the height of the interface at the horizontal position $x$ and the time $t$). We start the simulation from $h(x,t)=0$. To solve the EW equations we use the finite element method for both time ($t$) and space ($x$), whereas $h(x,t)$ is originally continuous. We use the first order (Euler) discretization method which shows clean enough scaling properties to yield the critical exponents required for the EW universality class and also for the case under study here. To model the noise (realized by RCP) we have to discretize also the $y$ axis, after which we have an $L_x\times L_y$ lattice in which the interface grows. The RCP noise is defined on the lattice, and to obtain it at a point on the interface at $x=i$ ($i$ being an integer), we calculate the integer part of $h$ at this point to detect the integer vertical coordinate, i.e. $j=\text{int}[h(i,t)]$ so that the interface falls into the site $(i,j)$ experiencing the random resisting force $V(i,j)$ to be inserted into Eq.~\ref{Eq:EW}. \\
RCP samples as the basic noise (as described in the previous section) are generated using the Eq.~\ref{Eq:random coulomb potential} with open boundary conditions. This has been already done in our previous works with exponents consistent with the RCP universality class~\cite{cheraghalizadeh2018gaussian,cheraghalizadeh2018gaussian2}. In this method one distributes charged (white noise) impurities with normal distribution over the lattice, and the Poisson equation is solved to obtain the potential filed, which serves as the quenched noise in the growth process of the interfaces. We generated over $10^5$ RCP samples, and simulated the motion of one interface for each sample using Eq.~\ref{Eq:EW}, i.e. we have generated $10^5$ interfaces for ensemble averaging. From now on, let us call the vertical direction ($y$ axis along which the interface grows) as the time direction, and the $x$ axis as the space direction. Generally we need samples with more extension along the time direction since the inteface needs more space to reach the steady state. In this work we considered samples with $L_y= 10L_x$, with $L_x\equiv L=32, 64, 128, 256, 512$ and $700$. The CPU time (12 cores with frequency $3.2$ GHz) was $\approx 1.04\times 10^7$s, i.e. about four months.\\

\begin{figure*}
	\begin{subfigure}{0.45\textwidth}\includegraphics[width=\textwidth]{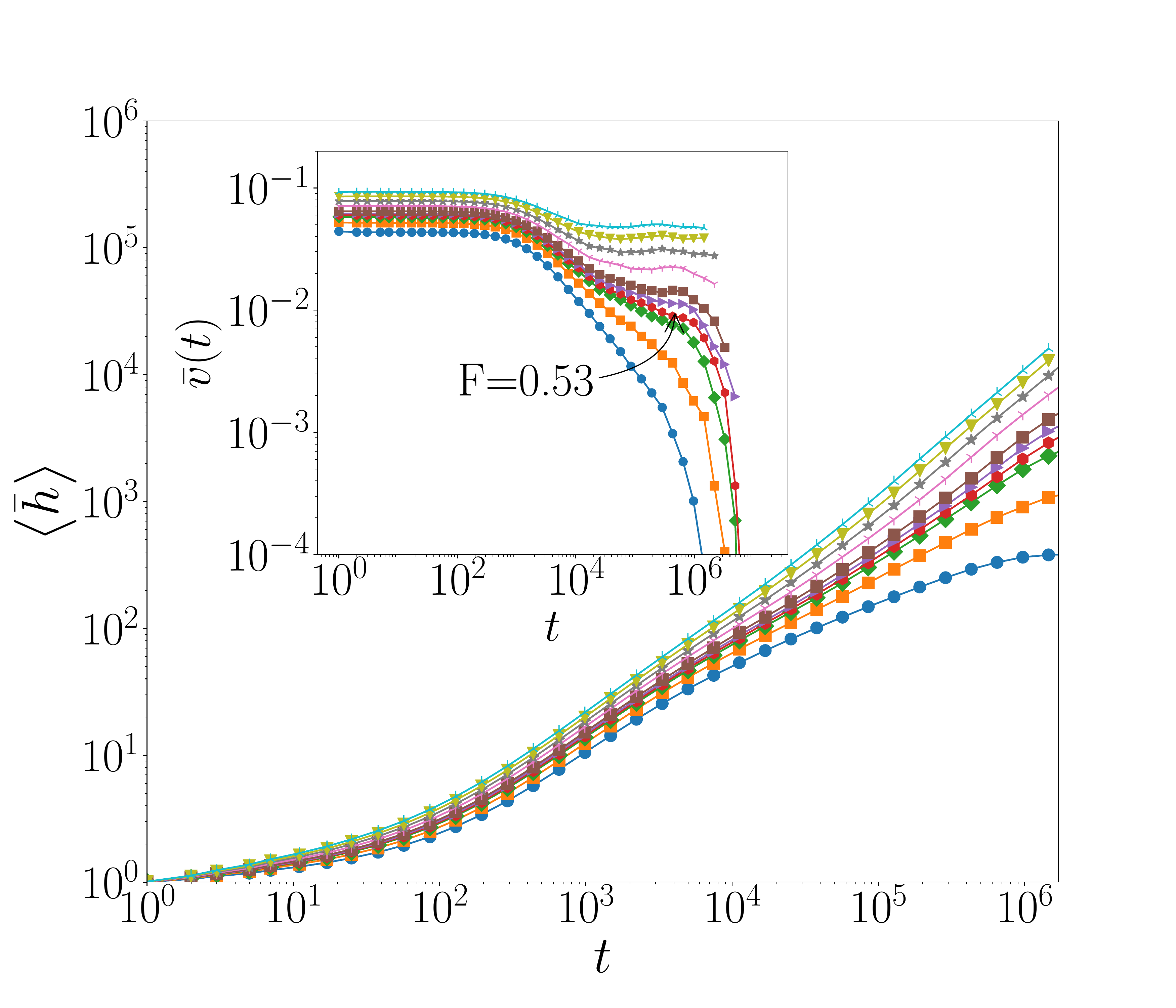}
		\caption{}
		\label{Fig:hvt}
	\end{subfigure}
	\begin{subfigure}{0.45\textwidth}\includegraphics[width=\textwidth]{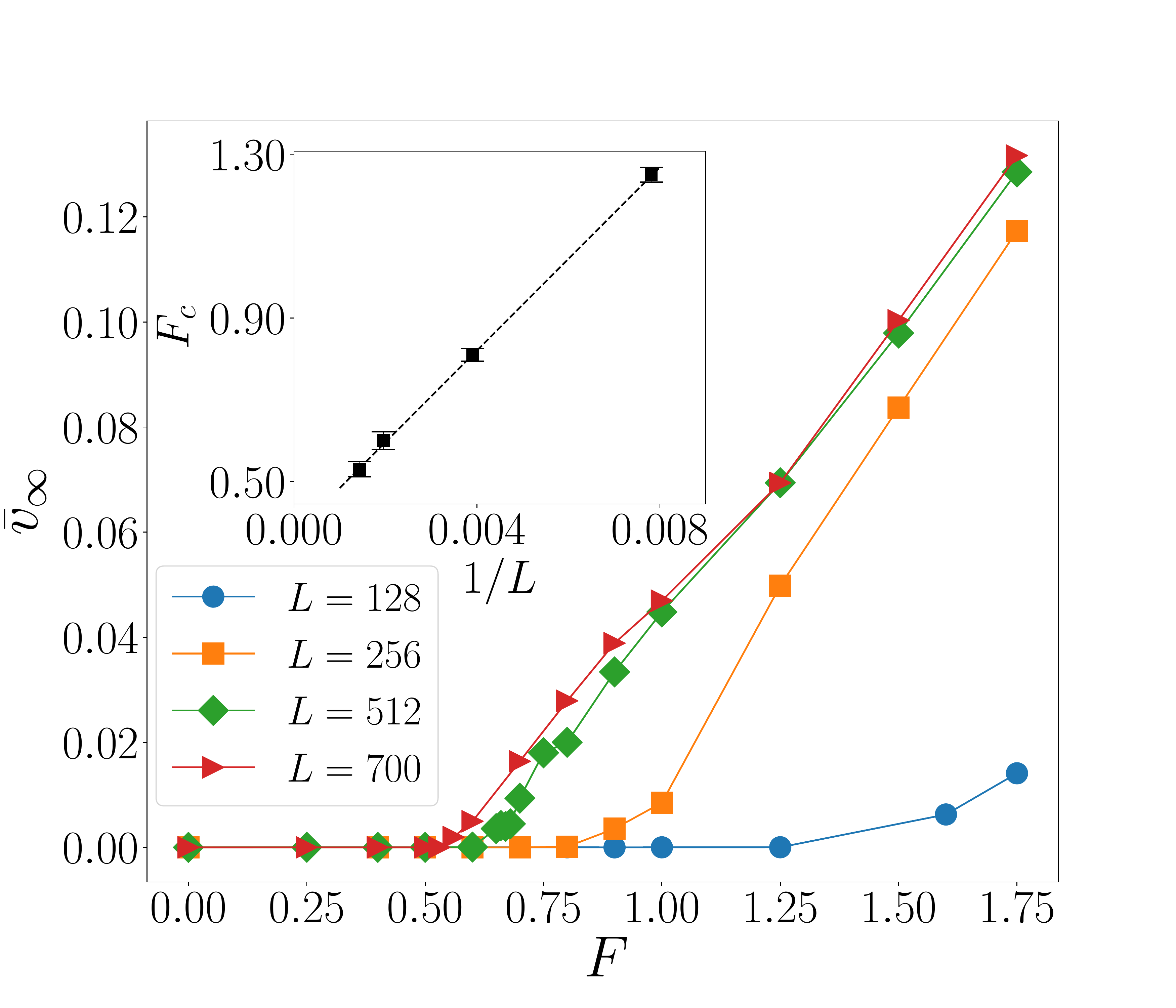}
		\caption{}
		\label{fig:Fc}
	\end{subfigure}
	\caption{(a) The plot of the average height $\left\langle\bar{h}\right\rangle $ versus time $t$ for various rates of the driving force $F=0.25,0.4,0.5,0.53,0.56,0.6,0.7,0.8,0.9,1$, in fact, the velocity strongly depends on the driving force $F$. Inset is log-log plots of velocity $v$ as function of time $t$ as obtained for different driving forces $F$ that shaw $F_{c}=0.53$ that measured for a system size $L=700$. after short time, for $F>F_{c}$ the interfaces grow at constant velocity, Therefore for  $F<F_{c}$ the interfaces become pinned. For all figures the results are obtained by starting with flat interfaces and averaged for 1000 realizations and by considering all the interfaces (pinned and moving) velocity of growing interface $v$ versus $F$, as measured for different lattice sizes $L$. we observe that all the interfaces become pinned below the critical point ($\bar{v}_{\infty}=0$). The pinning-depinning transition of the interfaces take places at critical values of driving forces of different lattice sizes, which can be estimated from left-hand side inset is plot critical values of driving forces $F_{c}$ vs  different lattice sizes $\frac{1}{L}$.}
	\label{fig:h-ave}
\end{figure*}

We analyze the time dependence of the ensemble average (denoted by $\left\langle \right\rangle $ for any observable) of the mean height of the interface, i.e. $\left\langle \bar{h}\right\rangle $ in terms of the driving force $F$ as is shown in Fig.~\ref{Fig:hvt}. Although for small times $\left\langle \bar{h}\right\rangle$ varies linearly with time for all $F$ values, for low enough $F$s at long time limit the graphs deviate from linearity in the log-log plot, i.e. bend downwards, showing a tendency of becoming constant in long enough times, i.e. $\bar{v}_{\infty}|_{\text{low} F}=0$ which is the characteristic of the pinned phase. For large $F$ values however, $\left\langle \bar{h}\right\rangle$ varies linearly with time in this limit, showing that the system is in the moving phase, and a depenning phase transition takes place in between. As is seen in the inset of Fig.~\ref{Fig:hvt}, $\bar{v}(t)$ is constant for early times for all $F$s, and crosses over to a new regime in the longer times which is power-law decay for low $F$s and nearly constant for large $F$s. For low $F$s, $\bar{v}(t)$ falls even faster for larger times, signaling that we are in the pinned phase according to the argument given in Eq.~\ref{Eq:dynamicalPowerLow} where an extra exponential decay factor is required (as mentioned in a line after Eq.~\ref{Eq:dynamicalPowerLow}). The graphs become almost constant at long times for $F>F_c$, showing that the system is in the moving phase. We have two possibilities to calculate $\bar{v}_{\infty}$ which is required for detecting the critical force $F_c$, which are: (1) consider the velocity at the largest time available in our data as an approximation of $\bar{v}_{\infty}$, and (2) extrapolate $\bar{v}(t)$ to find it. Since the extrapolation needs the precise fitting formula, the identification of which itself causes a large error (note that such a fitting formula is not clear for intermediate $Fs$), we considered the first strategy, and $F_c$ is estimated as the point were $\bar{v}_{\infty}$ becomes considerable (increase abruptly an order of magnitude) for the first time, see the Fig.~\ref{fig:Fc}. Our statistical analysis reveals that the critical force is $F_c=0.53\pm 0.02$ for $L=700$, which is identified by an arrow in the inset of Fig.~\ref{Fig:hvt} as the separator of the two phases. The velocity-time curve for $F=0.53$ does not however follow a power-law behavior as is evident in this figure, and instead the log-log plot of the orange curve ($F=0.4$) is linear for two decades. To understand this we considered the finite size dependence of $F_c$ which is presented in Fig.~\ref{fig:Fc}, inside which $F_c$ is plotted as a function of $1/L$ with a nice linear fit. This analysis shows that $F_c$ is extrapolated linearly to $0.37\pm 0.02$ as $L\rightarrow \infty$. Also we notice that $\bar{v}_{\infty}$ grows almost linearly with $F$ for $F>F_c$, i.e. $\theta\approx 1$ in Eq.~\ref{Eq:v-theta}.

\begin{figure*}
	\begin{subfigure}{0.45\textwidth}\includegraphics[width=\textwidth]{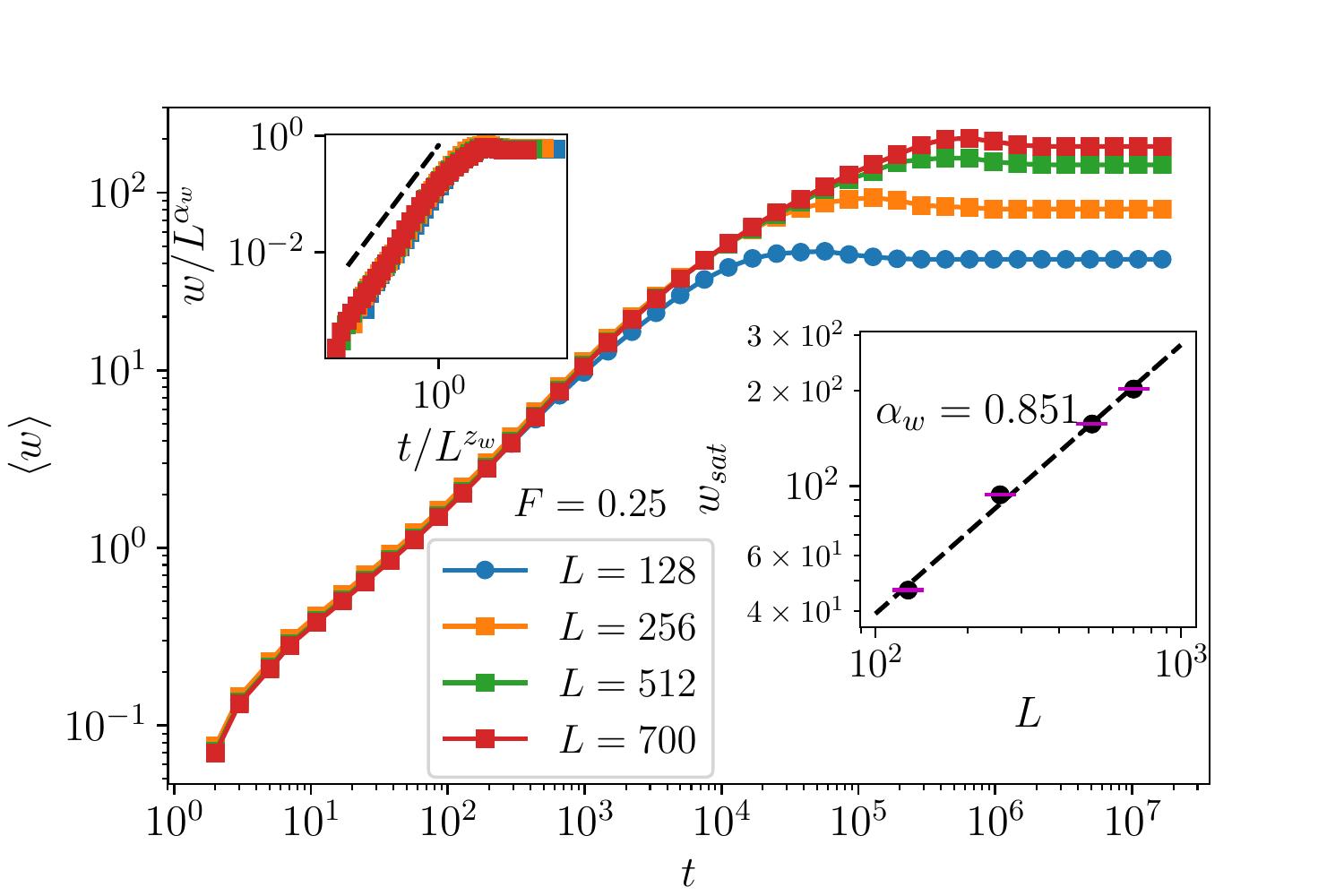}
		\caption{}
		\label{fig:wtL25}
	\end{subfigure}
	\begin{subfigure}{0.45\textwidth}\includegraphics[width=\textwidth]{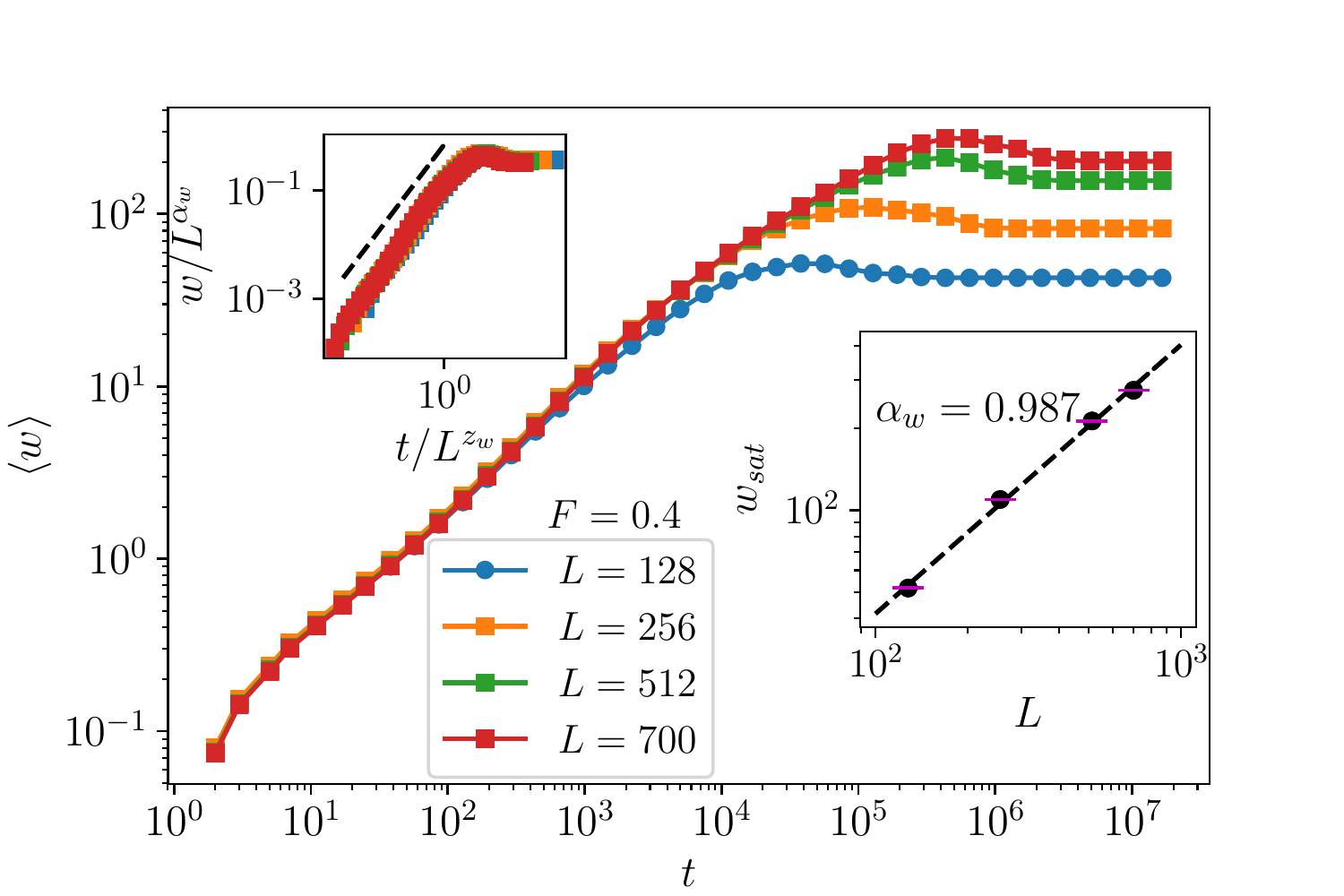}
		\caption{}
		\label{fig:wtL4}
	\end{subfigure}
	\begin{subfigure}{0.45\textwidth}\includegraphics[width=\textwidth]{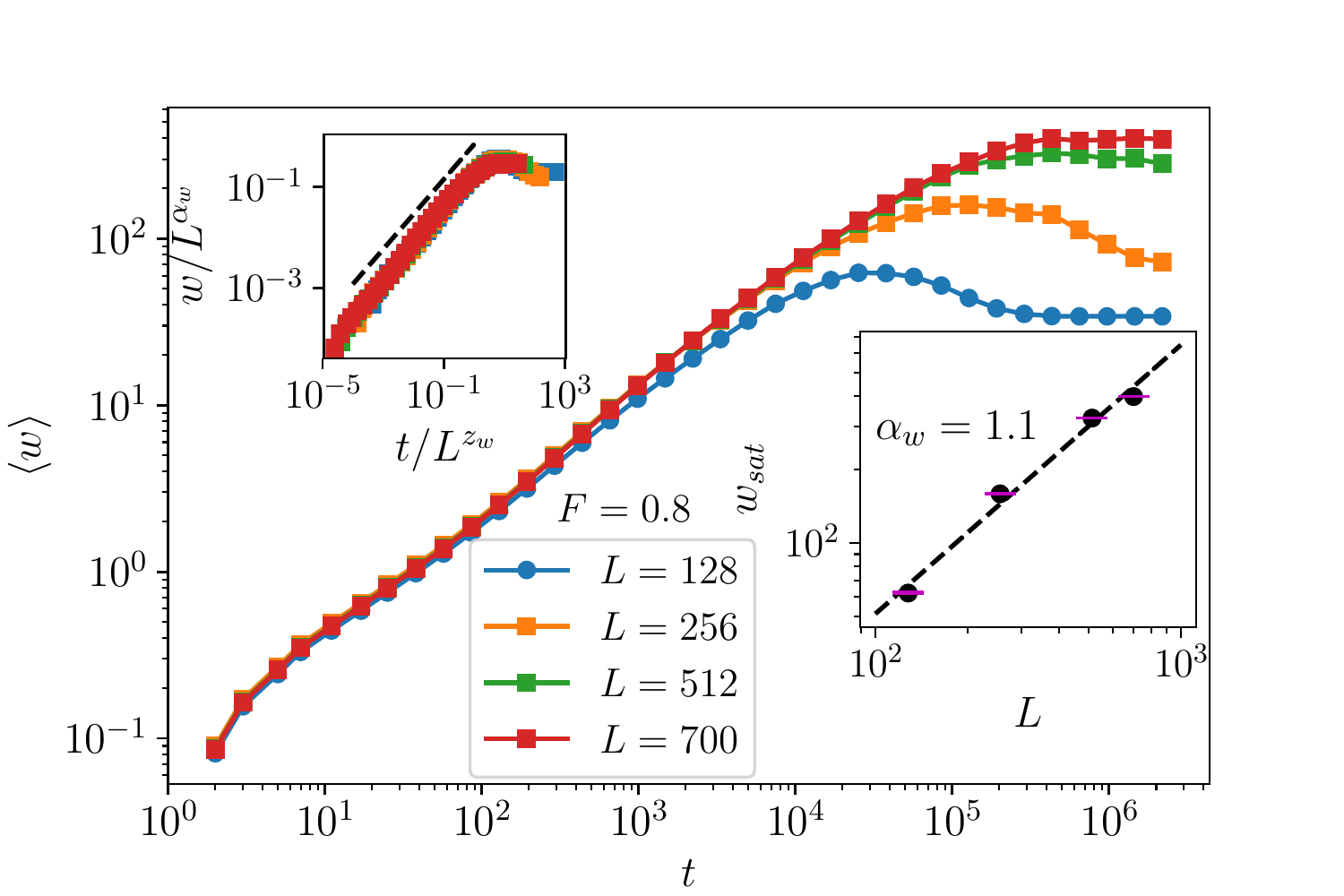}
		\caption{}
		\label{fig:wtL8}
	\end{subfigure}
	\begin{subfigure}{0.45\textwidth}\includegraphics[width=\textwidth]{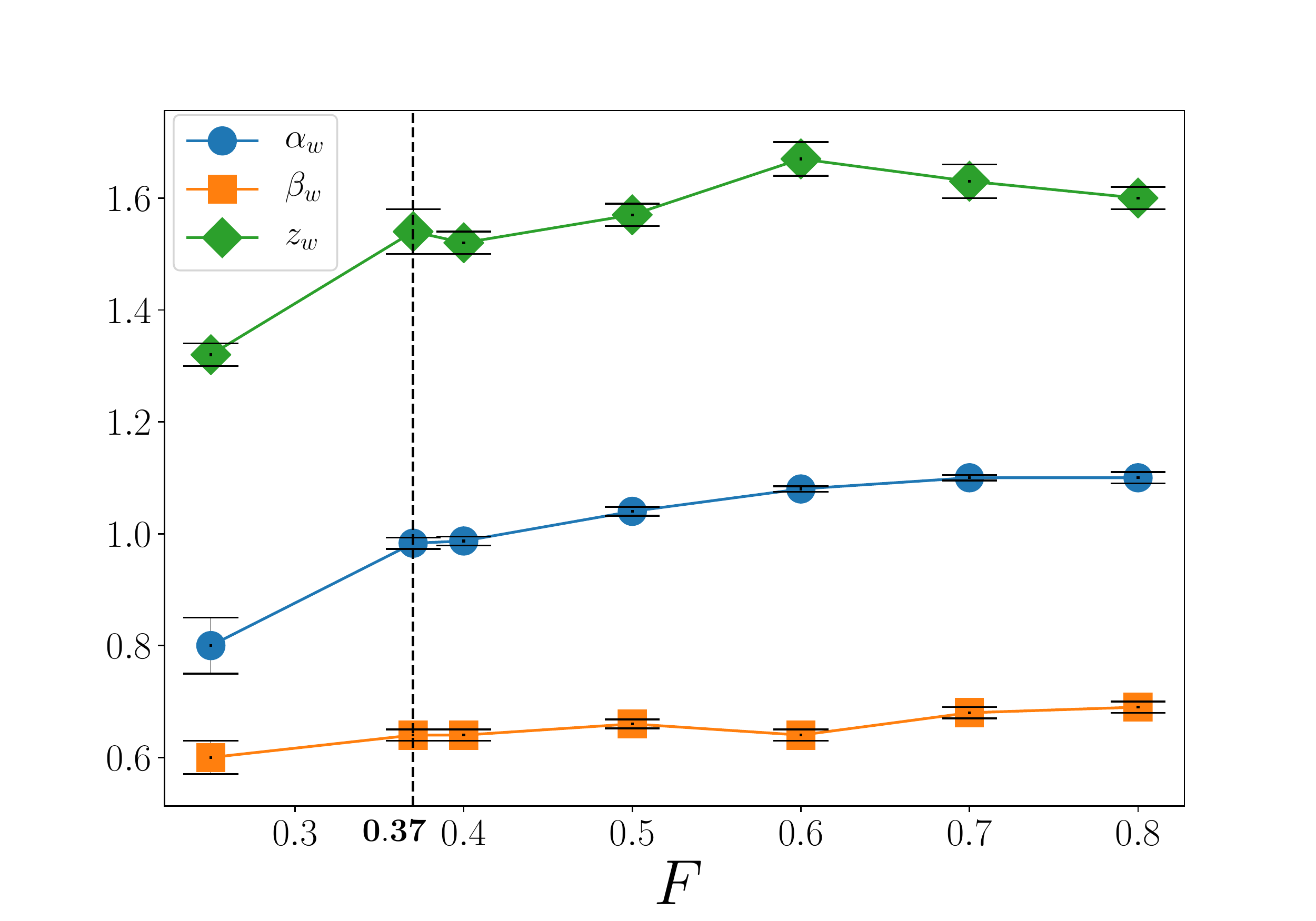}
		\caption{}
		\label{fig:Exponents}
	\end{subfigure}
	\caption{From the best fit of the data in the inset, we obtain: (a) For $F=0.25$, $\alpha_{w}=0.8 \pm 0.05$, $z_{w}=1.32\pm 0.02$ and $\beta_{w}=0.6 \pm 0.03$ (b) Log-Log plots of the average interface width $w$ vs time ($t$) for different size $ L$ in the driving force, $ F=0.4$, $\alpha_{w}=0.987 \pm 0.008$, $z_{w}=1.52 \pm 0.02$ and $\beta_{w}=0.64 \pm 0.01$ .(c) For $F=0.8$, $\alpha_{w}=1.10 \pm 0.01$, $z_{w}=1.6 \pm 0.02$ and $\beta_{w}=0.69 \pm 0.01$. (d) $\alpha_w$, $\beta_w$ and $z_w$ exponents in terms of $F$.}
	\label{fig:roughness}
\end{figure*}

As discussed in the arguments that led to Eq.~\ref{Eq:roughness-FSS}, we expect that the data collapse of the roughness of the interface gives us some new exponents, i.e. $\alpha_w$ and $z_w$. In the Figs~\ref{fig:wtL25},~\ref{fig:wtL4} and~\ref{fig:wtL8} we show the log-log plot of $\left\langle w\right\rangle $ versus time for different lattice sizes present, and also the data collapse analysis (the upper insets) for $F=0.25$, $0.4$ and $0.8$ respectively, each of which showing that the finite size scaling hypothesis Eq.~\ref{Eq:roughness-FSS}. The lower insets show the saturated roughness $w_{\text{sat}}$ in terms of $L$. The resulting exponents in terms of $F$ are shown in Fig.~\ref{fig:Exponents}. In contrast to $\beta_w$ which is nearly constant for all $F$ values, the exponents $\alpha_w$ and $z_w$ show a change at $F\simeq 0.37$. For $F\geqslant F_c\simeq 0.37$, $\alpha_w\simeq 1.05\pm 0.05$ and $ z_w\simeq 1.55\pm 0.05$, whereas for $F=0.25$ (bellow the transition point) we have $\alpha_w=0.8\pm 0.05$ and $z_w=1.32\pm 0.05$, which are different from the corresponding exponents $F>F_c$. For both of the phases the obtained exponents are different from the ones of the EW and KPZ classes. Therefore we see that Random Coulomb Potential correlated host changes the universality class of the driven interface. \\

\begin{table}
	\caption{Numerical results for exponents of interface velocity in different phases in near critical point. }
	\begin{tabular}{c | c c c }
		\hline quantity & $\alpha_{v}$ &  $z_{v}$ &  $\beta_{v}=\frac{\alpha_{v}}{z_{v}}$ \\
		\hline $F=0.25$ &     $0.15 \pm 0.04$  , & $0.42\pm 0.05$   ,  &   $0.34\pm 0.06$ \\
		\hline $F=0.4$ &    $0.14\pm 0.04$ &  , $0.42\pm 0.05$    ,& $0.32\pm 0.06$ \\
		\hline  $F=0.8$ &   $0.13\pm 0.03$ &  ,   $0.42\pm 0.05$    , & $0.31\pm 0.04$  \\
	\end{tabular}
	\label{tab:exponentsI}
\end{table}

\begin{figure*}
	\begin{subfigure}{0.45\textwidth}\includegraphics[width=\textwidth]{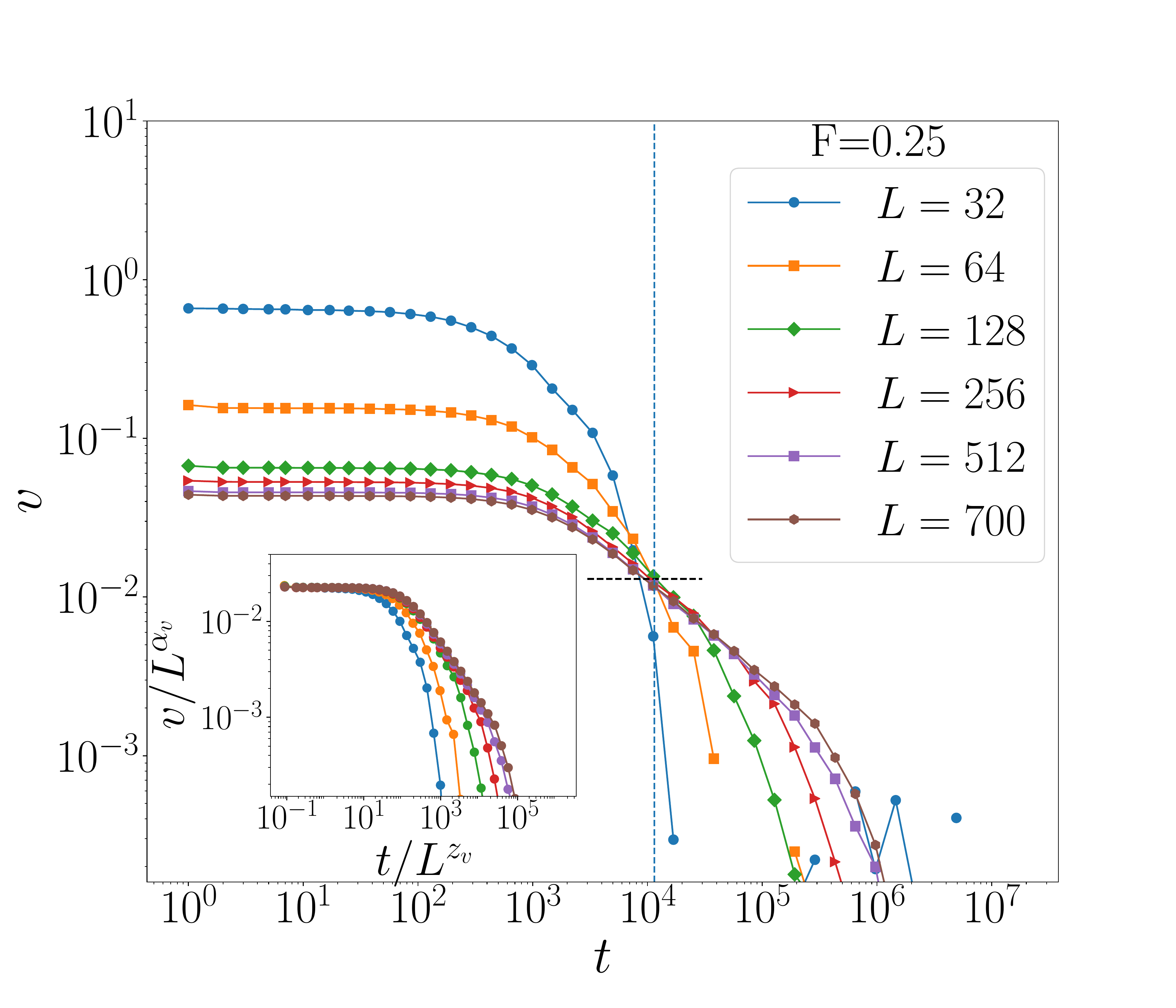}
		\caption{}
		\label{Fig:collapse25}
	\end{subfigure}
	\begin{subfigure}{0.45\textwidth}\includegraphics[width=\textwidth]{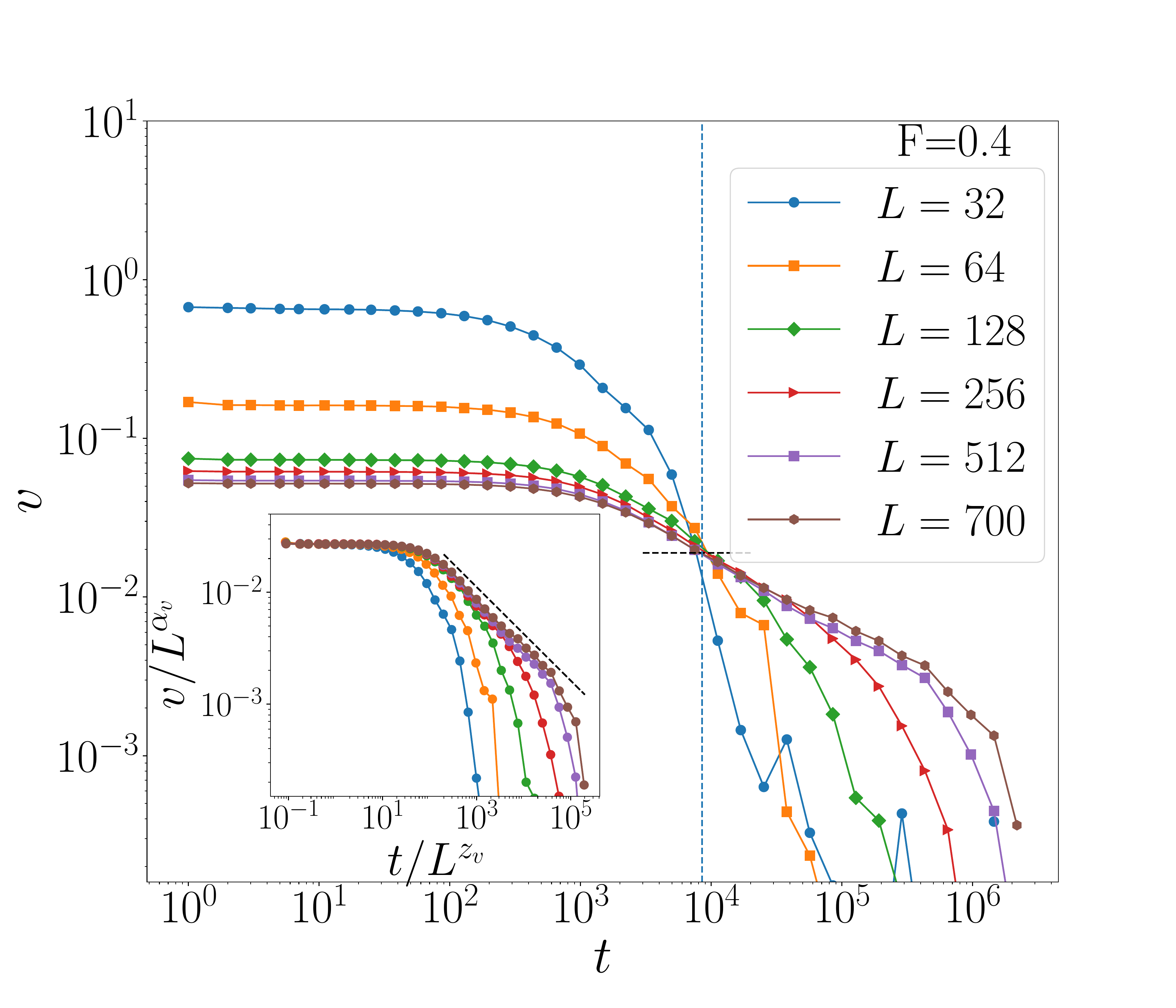}
		\caption{}
		\label{Fig:collapse4}
	\end{subfigure}
	\begin{subfigure}{0.45\textwidth}\includegraphics[width=\textwidth]{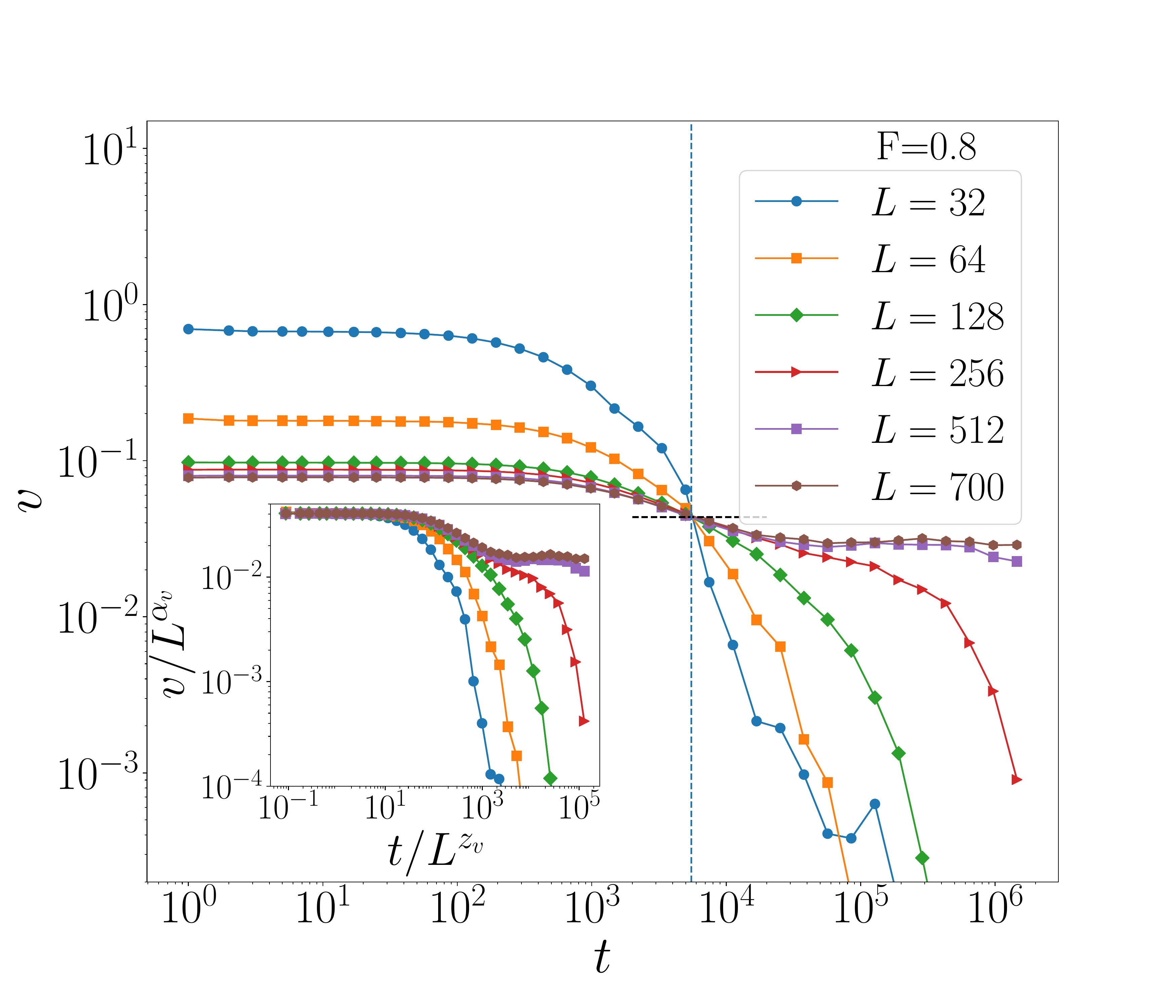}
		\caption{}
		\label{Fig:collapse8}
	\end{subfigure}
	\begin{subfigure}{0.45\textwidth}\includegraphics[width=\textwidth]{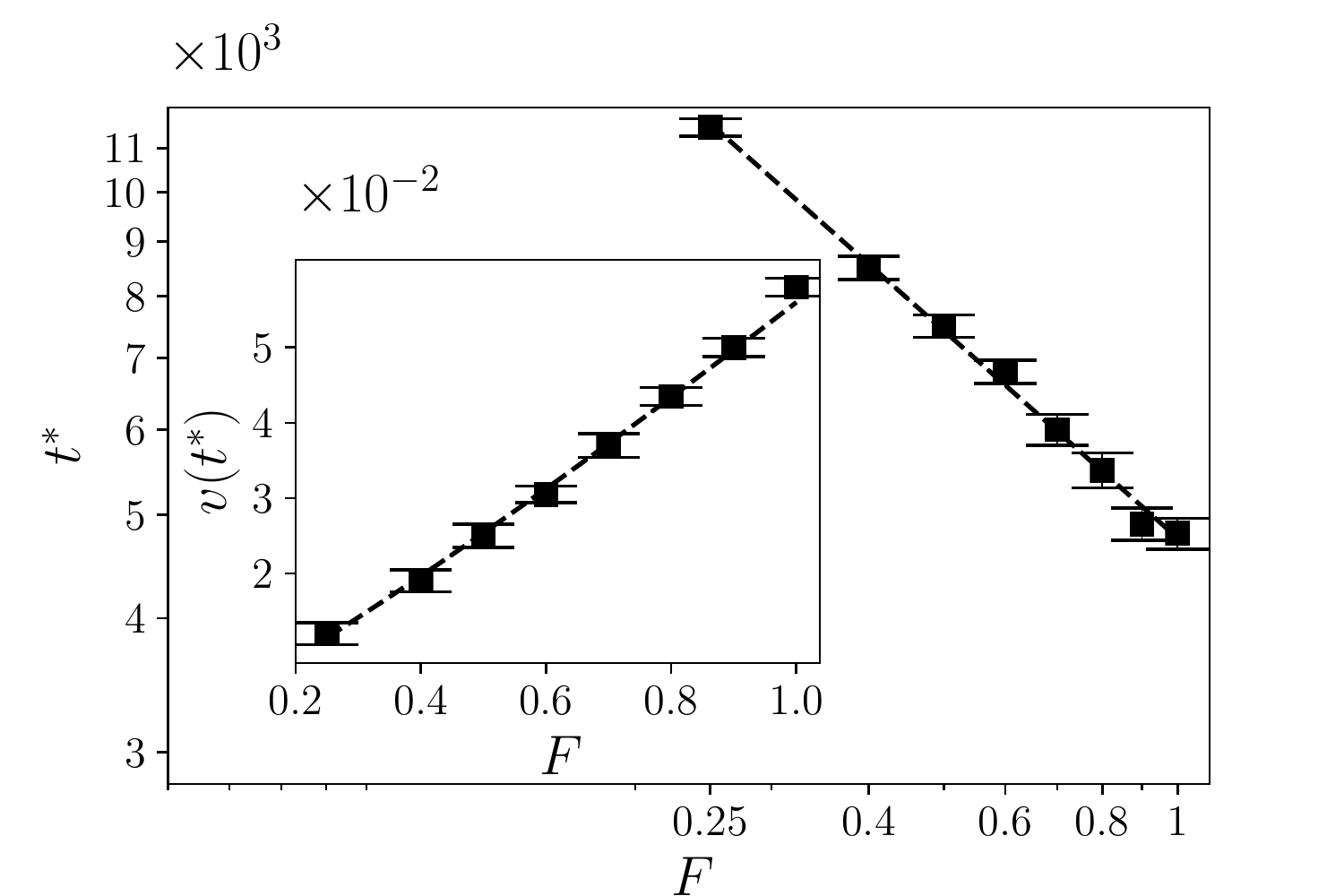}
		\caption{}
		\label{fig:VstarF}
	\end{subfigure}
	\caption{(a) log-log plots of $v$ versus $t$ at the driving force $F=0.25$, as measured for different lattice sizes $L$  ( all sizes are in pinned phase). From the best fit of the data in the inset. (b) log-log plots of $v$ versus $t$ For  $F=0.4$. (c) log-log plots of $v$ versus $t$ For  $F=0.8$, for a system size $L=700$. Numerical results for exponents of interface velocity in different phases are given in Table \ref{tab:exponentsI}.(d) log-log plot $t^{*}$ time in the crossover point versus $F$ and  the inset is log-log plot $v(t^{*})$ versus $F$. }
	\label{fig:dist1}
\end{figure*}

Actually $\bar{v}_F(t,L)$ deviates from the Eq.~\ref{Eq:dynamicalPowerLow}. To show this let us consider the $L$-dependence of the velocity curves in each phase, i.e. Fig~\ref{fig:dist1} where the results for $F=0.25$, $F=0.4$ and $F=0.8$ (Figs.~\ref{Fig:collapse25},~\ref{Fig:collapse4} and~\ref{Fig:collapse8} respectively) are shown. Interestingly we see that in all cases, the graphs cross each other in an almost \textit{single point}, denoted by $t^*$ which we interpret as the crossover point from pinned to moving phase. The slopes of the graphs at $t^*$ are definitely $L$-dependent. We see from Fig.~\ref{fig:VstarF} that $t^*\sim F^{-\eta}$, where $\eta=0.64\pm 0.04$. Also we observe that (Fig.~\ref{fig:VstarF}) $v(t^*)$ shows power-law dependence on $t^*$ and $F$ consistently, i.e. 
\begin{equation}
v^*\sim {t^*}^{-1.77\pm 0.1},\ \text{and}\ v^*\sim F^{1.14\pm 0.05},
\label{Eq:vstar}
\end{equation}
which is consistent with the amount of $\eta$ that we found.\\
Before exploring the properties of $t^*$ in more details, let us apply the data collapse analysis based on single scaling only the first part of the graphs ($t\ll t^*$), i.e. the following relation
\begin{equation}
v_{t\ll t^*}(t)=L^{\alpha_{v}^{(1)}}F_{v_1}\left(\frac{t}{L^{z_{v}^{(1)}}} \right)
\label{Eq:FirstPartCollapse}
\end{equation}
where $F_{v_1}$ is a function with $\lim_{x\rightarrow 0}F_{v_1}\left(x \right)=const$. $v_1$ represents that the analysis applies for the velocities only in the early times. This analysis, which one fits only the early parts of the graphs, is shown in the lower insets, the results of which are shown in the table (\ref{tab:exponentsI}). We see that the exponents do not vary with $F$. \\
\begin{figure*}
	\begin{subfigure}{0.45\textwidth}\includegraphics[width=\textwidth]{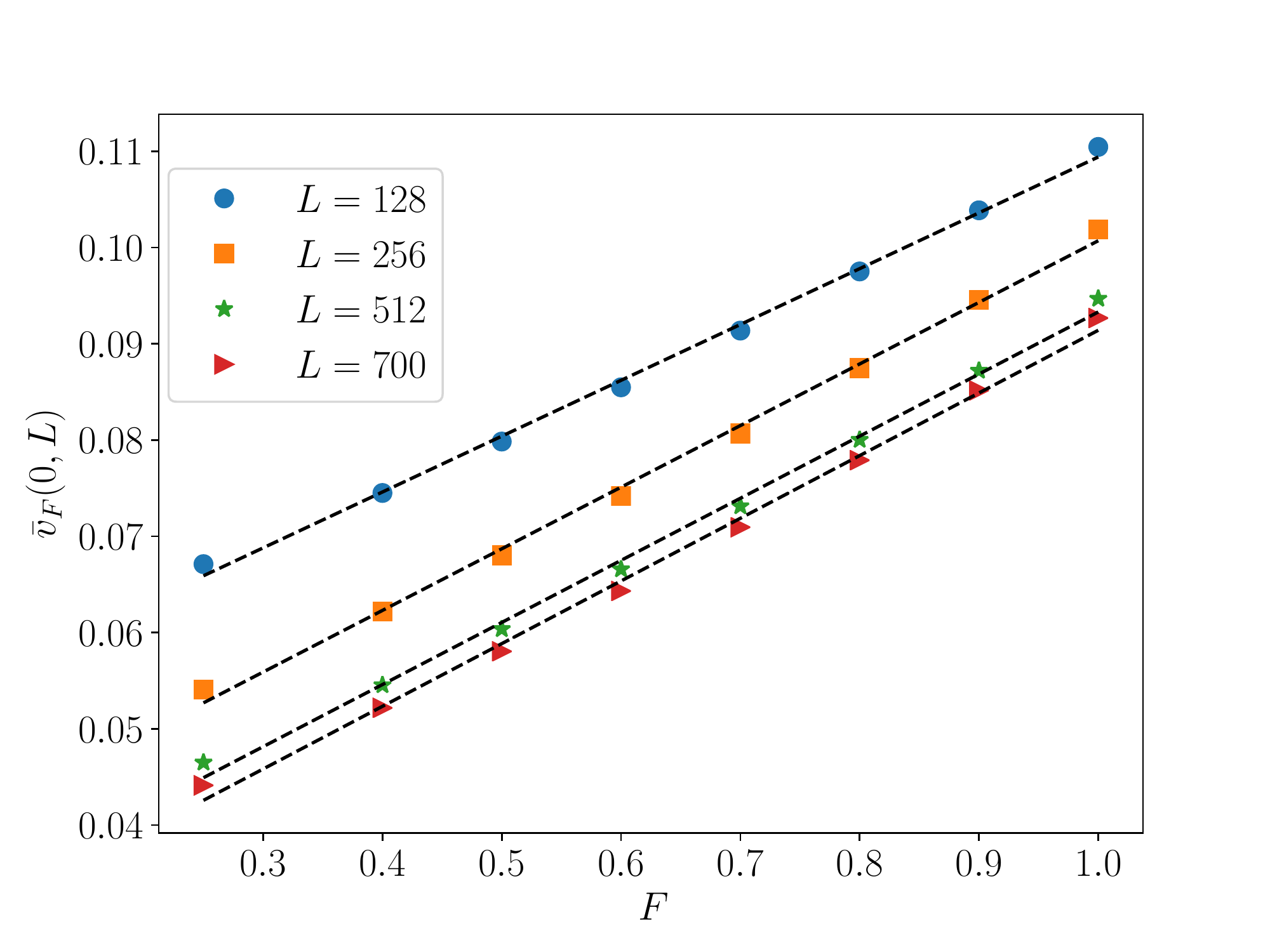}
		\caption{}
		\label{fig:vF}
	\end{subfigure}
	\begin{subfigure}{0.45\textwidth}\includegraphics[width=\textwidth]{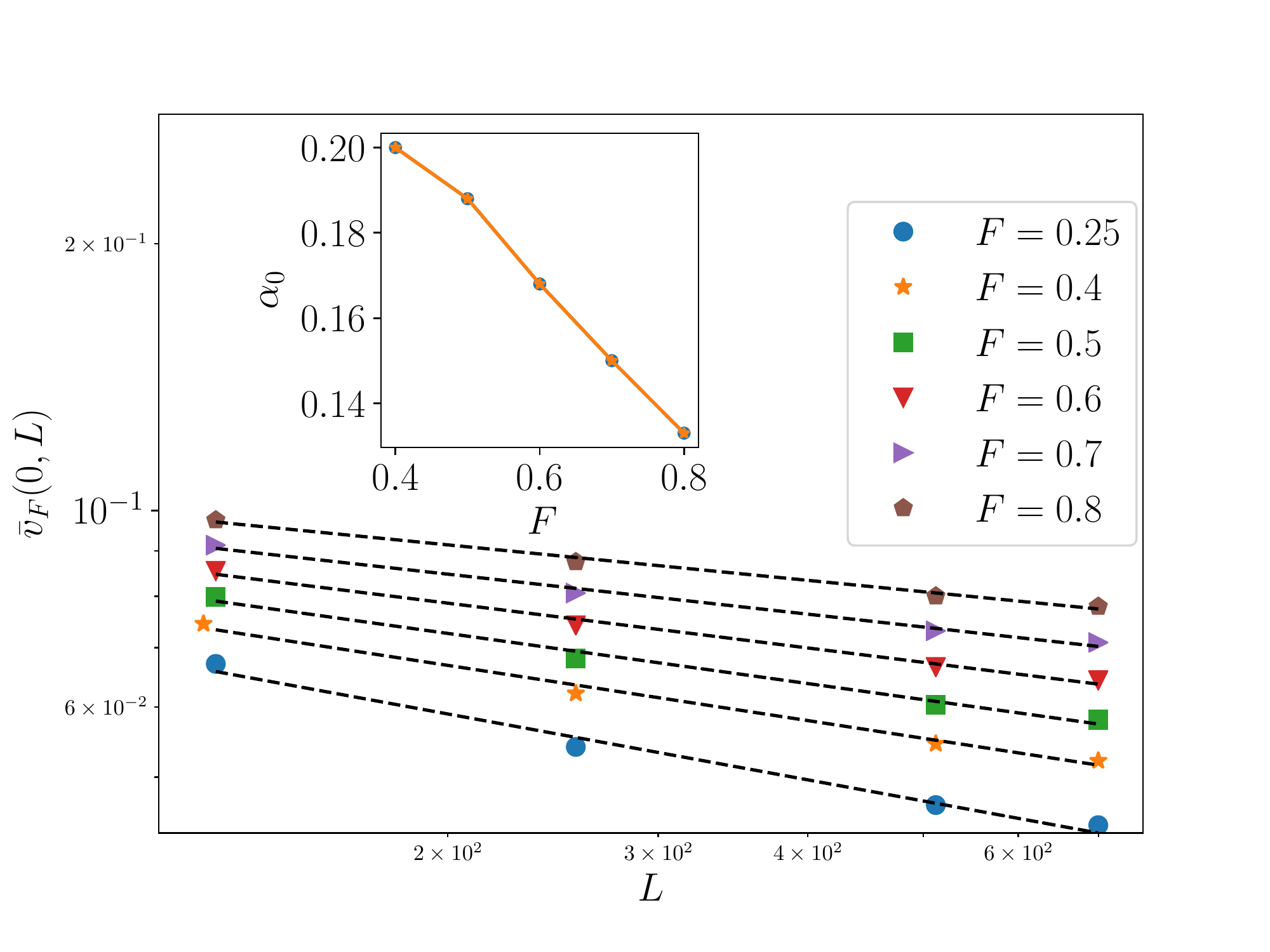}
		\caption{}
		\label{fig:vL}
	\end{subfigure}
	\caption{(a) The log-log plot of $\bar{v}_F(0,L)$ in terms of (a) $F$, and (b) $L$ with the exponents shown in the insets.}
	\label{fig:vF0}
\end{figure*}

\begin{figure}
	\centerline{\includegraphics[scale=.6]{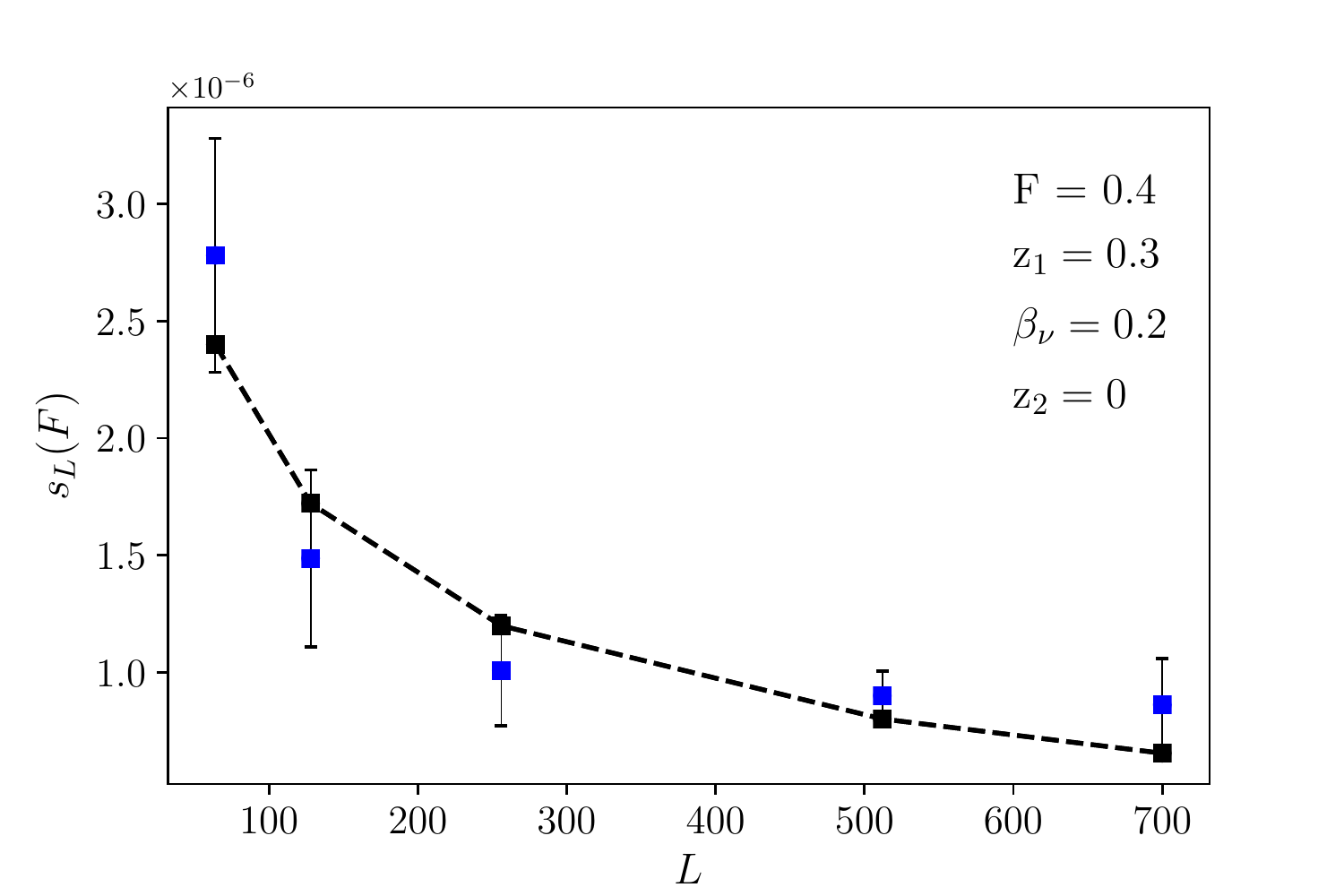}}
	\caption{The dependence of $s_{L}(F)$ versus $L$.}
	\label{fig:m1}
\end{figure}
It is important to note however that all parts of the velocity are not fitted with this relation, i.e. the portions in longer times are different. The most optimal way out of this inconsistency is to consider scaling relation with \textit{two variables}. To understand this, let us recall that there are two relevant spatial scales, namely the system size $L$ and the correlation length $\xi_F$. The later behaves like $\xi_F^{\infty}\equiv\xi_F^{L\rightarrow \infty}\sim (F-F_c)^{-\nu}$ in the thermodynamic limit close to the continuous transition points. This, alongside the fact that $\bar{v}_{F}(t,L)$ is $L-$independent at $t^*$ implies that the velocity is in the form 
\begin{equation}
\bar{v}_{F}(t,L)=L^{-\alpha_v}G\left(\frac{t-t^*}{L^{z_1}},\frac{t}{L^{\beta_v}{\xi_F^{\infty}}^{z_2}} \right)
\label{Eq:velocity}
\end{equation}
where $\alpha_v$, $\beta_v$, $z_1$ and $z_2$ are scaling exponents. We propose (and confirm later) that $G(x,y)$ asymptotically behaves like
\begin{equation}
G(0,y)\sim y^{-\alpha_v/\beta_v},\ G(-x,0)\sim a'x^{-\gamma}+b
\end{equation}
where $\gamma$ is a new exponent, and $a'$ and $b$ are some non-universal constants. Therefore, at $t=t^*$ we have
\begin{equation}
	v_F(t^*,L)\propto \left( \xi_F^{\infty}\right)^{\alpha_v z_2/\beta_v}{t^*}^{-\alpha_v/\beta_v} \ (\text{independent of}\ L)
\end{equation}
and also
\begin{equation}
	\lim_{t\rightarrow 0}v_F(t,L)=a'L^{-\alpha_v+\gamma z_1}{t^*}^{-\gamma}+bL^{-\alpha_v}
\end{equation}
To fix these exponents, we analyzed $\bar{v}_F(0,L)$ in terms of $L$ for fixed $F$ (and consequently fixed $t^*$), the results of which is shown in Fig~\ref{fig:vF0}. Figure~\ref{fig:vF} shows the dependence on $F$, which reveals that $\bar{v}_F(0,L)$ is linear with respect to $F$. Therefore, recalling that $t^*\propto F^{-\eta}$, one obtains $\gamma=\frac{1}{\eta}=1.56\pm 0.09$. This illustrates that
\begin{equation}
\lim_{t\rightarrow 0}\bar{v}_F(t,L)=m_L^{(1)}F+m_L^{(2)}
\end{equation}
where $m_L^{(1)}=aL^{-\alpha_v+\frac{z_1}{\eta}}$ and $m_L^{(2)}=bL^{-\alpha_v}$, and $a$ is a non-universal constant. The dependence of these slopes on $L$ is shown in~\ref{fig:vL}, giving $\alpha_v=0.39\pm 0.09$ and $z_1=0.29\pm 0.07$. \\

The only remaining exponents are $z_2$ and $\beta_v$, where are obtained by analyzing $v^*$. Since $v^*\equiv \bar{v}_{F}(t^*,L)$ doesn't vanish, nor become divergent at $F_c$ that we have found above (doesn't depend on $\xi_F^{\infty}$), we conclude that $z_2=0$, giving rise to
\begin{equation}
	v^*\sim {t^*}^{- \frac{\alpha_v}{\beta_v}}.
\end{equation}
Using Eq.~\ref{Eq:vstar} we find that $\beta_v=0.22\pm 0.07$. These exponents are shown in table~\ref{tab:exponents2}.\\

\begin{table}
\caption{The exponents of the velocity, given in Eq.~\ref{Eq:velocity}.}
	\begin{tabular}{c c c c c c }
		\hline  $\alpha$ & $\beta$ & $\gamma$ & $z_1$ & $z_2$& $\eta$ \\
		\hline $0.39(9)$ & $0.22(7)$ & $1.56(9)$ & $0.29(7)$ & $0$ & $0.64(4)$\\
		\hline
	\end{tabular}

	\label{tab:exponents2}
\end{table}

As an important check for the validity of Eq.~\ref{Eq:velocity}, we study the slope of $v_F$ in the vicinity of $t^*$. Consider two variables $x\equiv L^{-z_1}\left(t-t^*\right)$ and $y\equiv t\xi_F^{-z_2}L^{-\beta_v} $ so that $(x^*,y^*)\equiv \left(0,\frac{t^*}{L^{\beta_v}\xi_F^{z_2}} \right) $ is the crossover point, so that (noting from the above that $t^*$ is only $F$-dependent) $\bar{v}_F(x^*,y^*)$ is apparently $L$-independent. One can expand $\bar{v}_F$ in the vicinity of $x^*$ and $y^*$, giving rise to
\begin{equation}
\bar{v}_{F}(t,L)=v^* +(t-t^*)\left( \frac{A_F(L)}{L^{z_1}}+\frac{B_F(L)}{L^{\beta_v}\xi_F^{z_2}}\right) +...
\end{equation}
where $A_L(F)\equiv \partial_xG(x^*,y^*)=A\left(\frac{t^*}{L^{\beta_v}\xi_F^{z_2}} \right)$ and $B_L(F)\equiv \partial_yG(x^*,y^*)=B\left(\frac{t^*}{L^{\beta_v}\xi_F^{z_2}} \right)$ are the expansion coefficients. We then have 
\begin{equation}
\bar{v}_{F}(t,L)-v^*\propto s_L(F)(t-t^*).
\label{Eq:slopes}
\end{equation}
where 
\begin{equation}
s_L(F)=\frac{A_L(F)}{L^{z_1}}+\frac{B_L(F)}{L^{\beta_v}\xi_F^{z_2}}.
\end{equation}
The fact that $A_L(F)$ and $B_L(F)$ are $L$ and $F$ dependent, makes this analysis hard to do. For $F\simeq F_c$ however we found that these dependencies are negligible, and the obtained exponents are very close to the ones found above. The result is shown in Fig.~\ref{fig:m1}, in which the exponents are shown, i.e. $\beta_v=0.2\pm 0.03$ and $z_1=0.3 \pm 0.04$, which are consistent with the fittings just found in our previous analysis. From this study we infer that there is a crossover from $L<L^*$ to $L>L^*$, where $L^*=\left(A/B \right)^{\frac{1}{z_1-\beta}}\xi_F^{\frac{z_2}{z_1-\beta}} $. More precisely for $L>L^*$ we expect that the slope is given by $BL^{-\beta}\xi_F^{-z_2}$, whereas for $L<L^*$ it is $AL^{-z_1}$.\\

\begin{figure*}
	\begin{subfigure}{0.45\textwidth}\includegraphics[width=\textwidth]{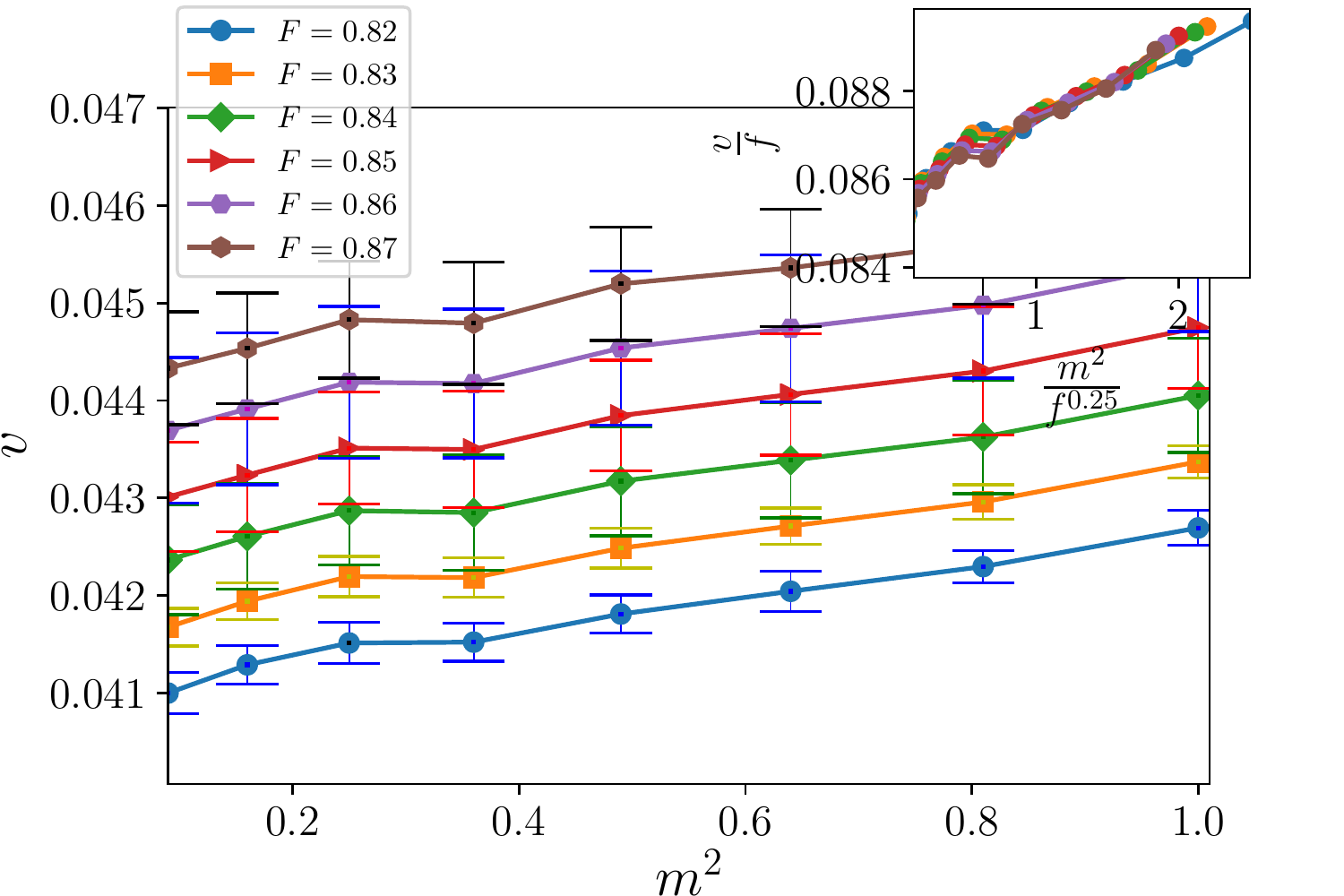}
		\caption{}
		\label{fig:slope1}
	\end{subfigure}
	\begin{subfigure}{0.45\textwidth}\includegraphics[width=\textwidth]{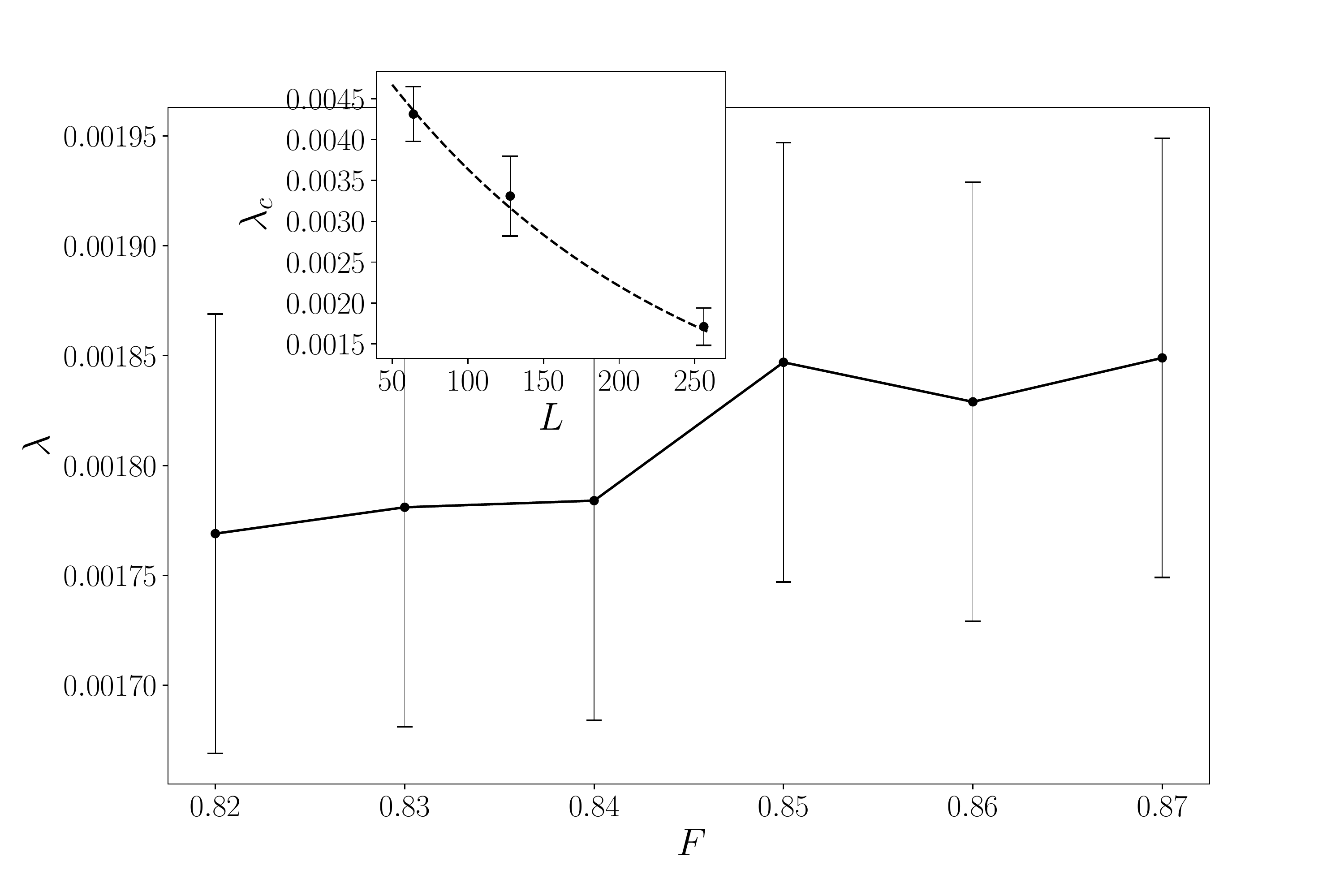}
		\caption{}
		\label{fig:slope2}
	\end{subfigure}
	\caption{(a) The dependence of $v_F$ of the tilted set up in terms of $m^2$. Inset: data collapse analysis of the main panel. (b) The slope of $v_F-m^2$ relation in terms of $F$, i.e. $\lambda(F)$. Inset shows $\lambda_c\equiv \lambda(F_c)$ in terms of the system size $L$, which decreases with $L$, extrapolating to zero as $L\rightarrow\infty$.}
	\label{fig:slope}
\end{figure*}

Now let us consider the velocities in the tilted setup, i.e. Eq.~\ref{Eq:tilted} using of which one can determine the universality classes. Figure~\ref{fig:slope} presents the results (the average velocity $v$) of the set up in which the initial interface is tilted with a slope $m$ for various driving forces, and for a system size $L=700$. This figure shows that $v$ varies almost linearly with $m^2$ (Fig.~\ref{fig:slope1}), especially in large $m$ values which confirms the relation~\ref{Eq:tilted}. We use the data collapse technique using the relation~\ref{Eq:tilted2}, which is shown in the inset of Fig.~\ref{fig:slope1}. With a simple re-scaling $\frac{v}{f}$ and $\frac{m^2}{f^{\frac{1}{4}}}$ we see that the curves fit to each other. It is seen that $\theta$ is almost 1, and $1-\phi=0.25$, giving rise to $\phi\approx \frac{3}{4}$. This shows that the $\lambda$ term does not diverge in the thermodynamic limit as the KPZ universality class. The $F$-dependence of $\lambda$ is depicted in Fig.~\ref{fig:slope2} in the vicinity of $F_c\approx 0.81$ for $L=256$, from which we see a nearly constant $\lambda$. For the $F_c$ this slope is a monotonic decreasing function of $L$, and extrapolates to zero in the $L\rightarrow\infty$ limit. This analysis shows again that the non-linearity term (which is responsible for lateral growth in KPZ) is not present in our model.

\section{Concluding Remarks}
In this paper we considered the effect of correlation in the (quenched) noise on a $L\times 10L$ square lattice to the depinning transition, and observed that it has a nontrivial effect on the motion and the morphology of an interface. To this end we considered the dynamics of interfaces which are described by quenched Edwards-Wilkinson (QEW) driven by the force $F$ on top of a lattice in which the noise results from a random coulomb potential noise (which corresponds to two-dimensional Edwards-Wilkinson model in the stationary regime). The interface is shown to be pinned by disorder if the driving force is small, and there is a critical force, shown by $F_c$ where the interfaces are critically pinned, so that for the forces just above this critical force the interface advances by jumping from one pinning path to another with a velocity almost proportional to $F$(see Fig.~\ref{fig:VstarF} that shows the final velocity $v_{f}$).  In the vicinity of the transition point the velocity varies according to Eq.~\ref{Eq:v-theta} with $\theta\approx 1$. In the moving phase, i.e. very large driving forces ($F \gg F_{c}$), the velocity of the interfaces are proportional to the driving force (the interfaces grow with constant speed). The analysis of the roughness shows that the growth exponents of our model are $\alpha_w=1.05\pm 0.01$ and the dynamical exponent $z_w=1.55\pm 0.05$, showing that the system is in a new universality class which is different from both EW and KPZ. \\

We also developed a two-variable scaling analysis for the velocity which is based on the observation of a crossover point where the velocity becomes $L$-independent. The exponents of this analysis are reported in table~\ref{tab:exponents2}. Up to our knowledge this type of two-variable scaling relation has not been seen before in the depinning transitions. \\

\bibliography{refs}

\end{document}